\documentclass[useAMS,usenatbib]{mn2e}
\usepackage{aas_macros} 
\usepackage{epsfig}

\title[Oscillatory convective modes in red giants]
  {Oscillatory convective modes in red giants: a possible explanation of the long secondary periods}
\author[H. Saio, P.R. Wood, M. Takayama, and Y. Ita]
  {Hideyuki Saio,$^1$,  Peter R. Wood$^2$
  Masaki Takayama$^{1,3}$, and  Yoshifusa Ita$^1$  \\
  $^1$Astronomical Institute, Graduate School of Science, Tohoku University,
     Sendai, Miyagi 980-8578, Japan\\
  $^2$Research School of Astronomy and Astrophysics, Australian National University, Cotter Road, Weston Creek, ACT 2611, Australia\\
  $^3$Department of Astronomy, School of Science, The University of Tokyo, Bunkyo-ku, Tokyo 113-0033, Japan
 }

\pagerange{\pageref{firstpage}--\pageref{lastpage}} \pubyear{2015}

\begin{document}

\label{firstpage}

\maketitle

\begin{abstract}
We discuss properties of oscillatory convective modes in low-mass red giants,
and compare them with observed properties of 
the long secondary periods (LSPs) of semi-regular red giant variables.
Oscillatory convective modes are very nonadiabatic g$^{-}$ modes and
they are present in luminous stars, such as red giants with $\log
L/{\rm L}_\odot \ga 3$.  Finite amplitudes for these modes are
confined to the outermost nonadiabatic layers, where the radiative
energy flux is more important than the convective energy flux.  The
periods of oscillatory convection modes increase with luminosity, 
and the growth times are comparable to the oscillation periods.  
The LSPs of red giants in the Large Magellanic Cloud (LMC) are observed to lie
on a distinct period-luminosity sequence called sequence D.  This
sequence D period-luminosity relation is roughly consistent with the
predictions for dipole oscillatory convective modes in AGB models
if we adopt a mixing length of 1.2 pressure scale height ($\alpha = 1.2$).
However, the effective temperature of the red-giant sequence of the LMC 
is consistent to models with $\alpha=1.9$,  which  
predict periods too short by a factor of two.
\end{abstract}

\begin{keywords}
 convection -- stars: AGB and post-AGB -- stars: late-type -- stars: low mass -- stars: oscillations .
\end{keywords}

\section{Introduction}

Long period variables (LPVs) are red giant stars which are known to
obey several period-luminosity relations \citep[e.g., ][]{woo99,
kis03,ita04,sos04,sos07}.  Among these relations, the so-called
sequence D relation corresponds to the longest period for a given
luminosity, with periods ranging from about 200~d to 1,500~d.  This
period is actually a long secondary period (LSP) co-existing with a
primary period which is about 8 times shorter and which typically
falls on sequence B \citep[e.g., ][]{woo99}.  Although other
sequences (except for sequence E) are generally associated with radial
or nonradial pulsations \citep{woo99,dzi10,tak13,mos13,woo15}, the
origin of sequence D is unknown.  Since the discovery of sequence D by
\citet{woo99}, its properties and origin have been discussed widely
\citep[e.g.,][]{woo04,sos07,nic09}.  Various models have been
considered such as semi-detached binaries \citep{woo99,sos07},
rotating spots with dust formation \citep{tak15}, and radial and
nonradial pulsation \citep{woo99}.  However, none of them give a
consistent explanation for the LSP variations.

There is considerable evidence that radial pulsation is not the main
mechanism associated with the LSPs.  The observed radial velocity
variations of a few km~s$^{-1}$ \citep{woo04,nic09}, combined with
the long periods, would produce a large fractional radius change if
the LSP is caused by radial pulsation.  However, there is no direct
evidence, for example from changes in effective temperature, for a
large radius change \citep{nic09}.  In addition, no change in the
primary period of the star is observed as might be expected if the
radius is changing significantly with the LSP variation \citep{woo04}.
Finally, we note that the fundamental mode radial pulsation period of
models is much shorter than the LSP \citep[e.g.][]{woo99}.  All these
factors indicate that the LSPs are not caused by radial pulsation.

One previous suggestion has associated convection with LSPs in red
giants.  \citet{sto10} considered giant convection cells for the
origin of the LSPs. He argued that the turnover time of giant
convection cells is comparable with observed length of LSPs, and that
the observed radial velocity changes can be interpreted as convection
motion, which causes no radius change.  However, it is not clear why
the turnover time would be related to a distinct periodicity with a
light amplitude of up to one magnitude.

In this paper we discuss oscillatory convective modes in red giants.
They are oscillations confined to the outermost layers of the deep envelope convection.  
Dipole oscillatory convective modes have properties favorable for the cause of the LSPs,
such as periods longer than the fundamental radial mode and small temperature
variations.
We also note that the above mentioned difficulties in radial pulsations 
associated with the observed radial velocity variations of LSPs 
are not applicable for a dipole mode pulsation because it does not change 
mean stellar radius.

We discuss evolutionary models on which we performed pulsation
analysis in Section \ref{sec:model}. The properties of the oscillatory
convective modes are discussed
in Section \ref{sec:oscillation}. We make comparisons with observations
of LSP variability in Section \ref{sec:discussion}.

\section{Models}\label{sec:model}

\begin{figure}
\begin{center}
\epsfig{file=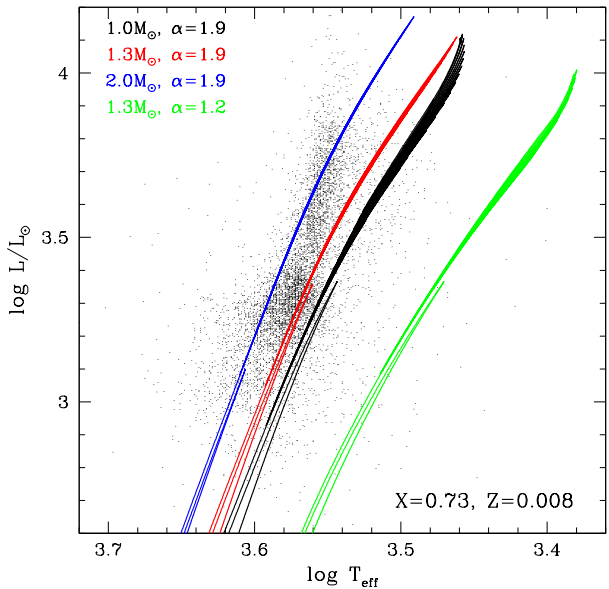,width=0.48\textwidth}
\end{center}
\caption{Selected evolutionary tracks computed by the MESA code for initial masses of 
$1.0~{\rm M}_\odot$, $1.3~{\rm M}_\odot$ and $2.0~{\rm M}_\odot$ with the mixing-length parameter, $\alpha = 1.9$ and $1.2$. The evolutions were started at a pre-main sequence model and calculated through the thermal pulsing AGB stage up to $\log L/{\rm L}_\odot \approx 4.0$.
Small dots are sequence D stars identified by \citet{sos09} (OGLE III).
}
\label{fig:hrd}
\end{figure}

Fig.~\ref{fig:hrd} is a HR diagram showing the positions 
of sequence D stars (small dots),  and some evolutionary tracks 
on the red-giant branch and asymptotic giant branch (AGB) for initial masses of
$1.0, ~1.3$,  and $2.0~{\rm M}_\odot$ with $\alpha = 1.9$ and $1.2$,  
where $\alpha$, a mixing-length
parameter, is defined as the mixing-length divided by the pressure scale-height.
Sequence D stars are adopted from OGLE III identification by \citep{sos09}.
The  luminosity and effective temperatures were obtained from
2MASS photometry data of $J$ and $K$ \citep{2mass} 
and the temperature and bolometric correction - color relations 
by \citet{hou00} and \citet{hou00b}, 
and a distance modulus of 18.54 mag \citep{kel06} of the Large Magellanic
Cloud (LMC). 
The 2MASS $J, K$ magnitudes were converted to the CIT system 
using the relation  derived by \citet{car01}.

Stellar evolution models were
obtained using the MESA (Modules for Experiments in Stellar
Astrophysics; version 7184) code \citep{pax13}. The hydrogen and metal
abundances adopted are $(X, Z) = (0.73, 0.008)$.  Models of various
initial masses ranging from $0.9~{\rm M}_\odot$ to $2.0~{\rm M}_\odot$
were computed with several values (between $1.0$ and $1.9$) of the
mixing-length parameter $\alpha$. 

The central helium flashes start to occur near the tip of the red giant branch (RGB) at
$\log L/{\rm L}_\odot \approx 3.3$.  After a few shell flashes, the
star transits to a steady central helium burning around $\log L/{\rm
  L}_\odot \sim 2$. After central helium exhaustion, the evolution
along the asymptotic giant branch (AGB) starts.  Here helium shell
flashes (thermal pulses) occur in which the luminosity changes
cyclically while the average luminosity increases gradually. The
evolution calculations were stopped when each model reaches at $\log
L/{\rm L}_\odot \approx 4.0$.  
Wind mass loss rate of \citet{dej88} was included in the calculations.  

The majority of the sequence D stars are AGB stars, whose loci 
on the HR diagram look consistent with evolution models with $\alpha = 1.9$.

\section{Oscillatory convective modes in red giants}\label{sec:oscillation}

\citet{shi81} discovered that nonradial g$^-$ modes, whose frequencies
are purely imaginary (meaning monotonic growth or decay) in the
adiabatic condition, corresponding to the convective instability,
become oscillatory in extremely nonadiabatic condition.  We call them
oscillatory convective modes.  They are highly unstable (or strongly
excited) with growth time comparable with the period.  \citet{shi81}
discovered the oscillatory convective modes in studying nonradial
pulsations of high angular degrees $\ell \ga 10$ in luminous
$(10^5~{\rm L}_\odot)$ models hotter than the cepheid instability
strip.  Such high degree modes are not expected to be observable
because of cancellation on the surface.  Thirty years later,
\citet{sai11} found low
degree $\ell \la 2$ examples of these oscillatory convective modes 
in hot massive stars where
convection zones are associated with the Fe opacity peak at $T\sim 2\times10^5$
K.

\begin{figure}
\begin{center}
\epsfig{file=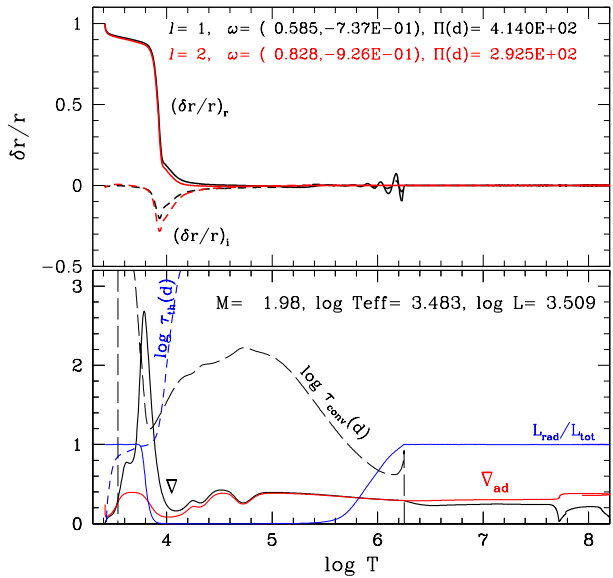,width=0.48\textwidth}
\end{center}
\caption{
{\bf Top panel:} Amplitude distributions of  typical oscillatory convective modes of $\ell = 1$ (black lines) and $\ell = 2$ (red lines) are shown as a function of $\log T$ in a model on the AGB branch for a initial mass of $2.0~{\rm M}_\odot$ with a mixing-length parameter of $\alpha=1.2$. Solid and dashed lines are, respectively, the real and imaginary parts of the normalized radial displacement $\delta r/r$. The real and imaginary parts of the normalized angular frequency $\omega$ and the period $\Pi$ in days for each mode are shown.  We adopt in this paper the convention that a pulsation mode is excited if the imaginary part is negative (i.e., $\omega_{\rm i} < 0$).
{\bf Bottom panel:}  Some physical variables in the AGB model are shown, where $\tau_{\rm conv}$ and $\tau_{\rm th}$ are, respectively, convective turn-over time and thermal time in units of days.
}
\label{fig:eigfunL1L2}
\end{figure}

In this paper we consider oscillatory convective modes in luminous red
giants having deep convection envelopes.  Frequencies and amplitude
distributions in the stellar interior were obtained by the method
described in \citet{sai80}, in which the Lagrangian perturbation for
the divergence of convective flux is neglected (i.e., $\delta
\nabla\cdot\boldmath{F}_{\rm conv}=0$).  The temporal variation is
expressed as $\exp(i\sigma t)$ with complex angular frequency
$\sigma$, so that an oscillation mode is excited (or overstable) if
$\sigma_i < 0$, where $\sigma_i$ is the imaginary part of $\sigma$.  
To represent the value of eigenfrequency, we use
normalized (non-dimensional) eigenfrequency $\omega$ defined as
\begin{equation}
\omega = {\sigma\sqrt{R^3\over GM}},
\label{eq:omega}
\end{equation}
where $R$, $M$, and $G$ are stellar radius and mass, and gravitational
constant, respectively i.e., the pulsation frequency is
normalized by the Kepler frequency at the stellar surface.

The top panel of Fig.~\ref{fig:eigfunL1L2} shows the amplitude
of the radial displacement $\delta r/r$ for oscillatory convective modes
of $\ell = 1$ (black lines) and 2 (red lines) in the interior of an
AGB model of an initial mass of $2.0~{\rm M}_\odot$ at $\log L/{\rm
  L}_\odot = 3.5$ with $\alpha=1.2$.  Solid and dashed lines are, respectively, real
and imaginary parts of the radial displacement.  Actual radial
displacement is given by the real part of
$\delta r e^{i\sigma t}Y_\ell^m(\theta,\phi)$ where
$Y_\ell^m(\theta,\phi)$ is  a spherical harmonic and $\theta$ and $\phi$ are
the co-latitude and azimuthal angle, respectively.

The amplitude distributions are remarkably similar for the $\ell =1$
and $2$ modes.  Finite amplitudes are mostly confined to the outer
layers where the superadiabatic temperature gradient $\nabla -
\nabla_{\rm ad}$ is large and the thermal time $\tau_{\rm th}$ is
shorter than the periods, indicating that the oscillatory convective modes are
very nonadiabatic.  Here, $\nabla$ and $\nabla_{\rm ad}$ are defined as
\begin{equation}
\nabla = {d\ln T\over d\ln P}, \quad \nabla_{\rm ad} = \left(\partial\ln T\over\partial\ln P\right)_S,
\end{equation}
where $P$ and $S$ are the pressure and the specific entropy, respectively.
The thermal time $\tau_{\rm th}$ is defined as
\begin{equation}
\tau_{\rm th} = {(M-M_r)C_pT\over L_r},
\label{eq:thermal}
\end{equation}
where $M_r$ is the mass within the sphere of radius $r$ (i.e., $M-M_r$
is the mass above the sphere of radius $r$), $C_p$ is the specific heat
at constant pressure, and $L_r$ is the local luminosity at radius
$r$.

Also shown in the bottom panel of Fig.~\ref{fig:eigfunL1L2} is
the local convective turnover time defined as
\begin{equation}
\tau_{\rm conv} = {\alpha H_p\over v_{\rm conv}},
\end{equation}
where $H_p$ is the pressure scale height, and $v_{\rm conv}$ is
convective velocity obtained from the mixing-length theory.  Note that
$\tau_{\rm conv}$ decreases toward the inner boundary at $\log T
\approx 6.2$; this is because the fractional radius of the convective
inner boundary is very small ($r/R\approx 0.006$, while $M_r/M \approx
0.28$).  Fig.~\ref{fig:eigfunL1L2} indicates that in a considerable
fraction of the zone where the radial displacement has a large
amplitude, $\tau_{\rm conv}$ is much larger than the period and most
of the energy is carried by radiation (i.e., $L_{\rm rad}/L_{\rm tot}
\approx 1$). 
This means that our approximation neglecting the
divergence of convective flux is likely to be reasonably good.

The absolute value of the imaginary part of the non-dimensional
eigenfrequency $\omega $ is comparable with the real part, indicating that
the growth time is comparable with the pulsation period.  In
other words, the oscillatory convective modes are violently excited so
that they should be capable of causing semi-regular light/velocity variations
of the size observed \citep[e.g.,][]{woo99,sos04} during LSP variations.

\begin{figure}
\begin{center}
\epsfig{file=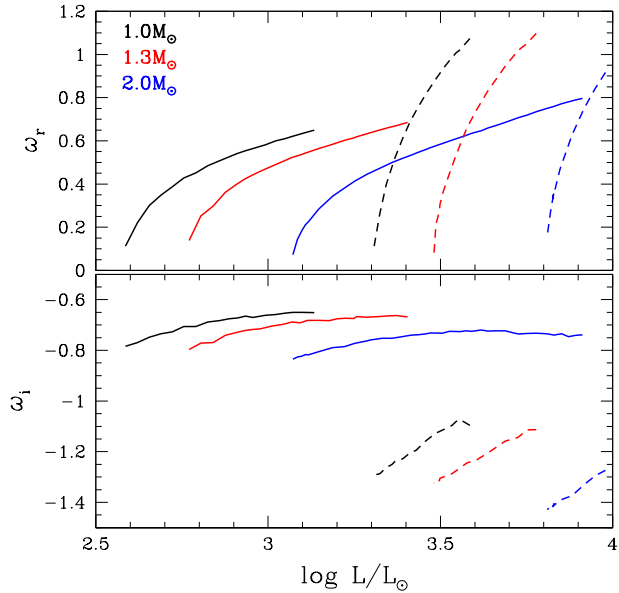,width=0.48\textwidth}
\caption{Real (upper panel) and imaginary (lower panel) of normalized eigenvalue
of oscillatory convective modes in  AGB (non-flashing) models of each mass
(black lines -- $1.0~{\rm M}_\odot$; red lines -- $1.3~{\rm M}_\odot$; 
blue lines -- $2.0~{\rm M}_\odot$). 
Solid and dashed lines are for models with $\alpha = 1.2$ and $1.9$, respectively.
}
\end{center}
\label{fig:lum_omega}
\end{figure}

Fig.~\ref{fig:lum_omega} shows the real part ($\omega_r$; upper panel) 
and imaginary part ($\omega_i$; lower panel) of normalized eigenfrequencies 
of oscillatory convective modes as a function of the luminosity of
AGB models of various masses and with mixing-length parameters
$\alpha = 1.2$ (solid lines) and $1.9$ (dashed lines).
(We have selected only non-flashing, i.e., slowly evolving,  AGB models.)
Generally, $\omega_r$ increases and $|\omega_i|$ decreases 
with luminosity, because non-adiabaticity
is stronger (or thermal time is shorter) in more luminous models.
Conversely, as luminosity decreases, $\omega_r$ decreases and become essentially
equal to zero at a certain luminosity for a given mass and $\alpha$, below which
the mode behaves as a monotonic g$^{-}$ mode.
This means that there is a minimum 
luminosity for the presence of oscillatory convective modes depending on the mass
and $\alpha$.
The minimum luminosity is higher for larger mass for a given $\alpha$, 
while it is higher for larger $\alpha$ for a given mass (Fig.~\ref{fig:lum_omega}).
These dependences can be understood using the thermal time given in 
eq.~(\ref{eq:thermal}) which should be inversely proportional to the strength
of nonadiabaticity.
For a larger mass model, a higher luminosity is needed 
to compensate a larger numerator, $M-M_r$, 
of eq.~(\ref{eq:thermal}); the same is true for a model with larger $\alpha$,  
because radius is smaller and hence mean density of the envelope is larger, 
which means that $M-M_r$ is larger for a given temperature.

As seen in Fig.~\ref{fig:lum_omega}, $\omega_r$ increases as luminosity increases
and become nearly equal to the absolute value of the imaginary part, $|\omega_i|$, 
where each line is terminated.
We have terminated the sequence because the amplitude in the radiative core 
becomes extremely large ($> 10^4$) and rapidly oscillating for a mode with  
$\omega_r \ga |\omega_i|$, although 
the amplitude distribution in the envelope hardly differs from that shown 
in Fig.~\ref{fig:eigfunL1L2}.
Such a huge amplitude in the core is  caused by a coupling with a high-order g-mode; 
i.e., the mode is a mixed mode consisting of an oscillatory convective mode 
in the outer envelope and high-order g mode in the core.  
It is not possible to resolve well such a large amplitude eigenfunction in the core 
(where $r/R < 10^{-4}$ ) with a few ten thousands mass shells we are using. 
In such a mode, it is expected that non-linear effects in the core would become 
significant before the surface amplitude becomes appreciable.
For this reason we consider only pure oscillatory convective modes with
$\omega_r \la |\omega_i|$ to be observable.

Nonradial pulsations in a red giant are known to have  mixed mode properties, 
with p-mode characteristics in the envelope and g-mode characteristics in the core \citep{dzi01}. 
The trapping of the p-mode in the envelope becomes very efficient as the luminosity
 increases as discussed by \citet{dup09}. 
However, the coupling between an oscillatory convective mode in the envelope and 
a core g-mode is different from the p- and g-mode coupling. In the case of  the 
oscillatory convective mode,  no isolation of the envelope oscillation from the core 
g-mode mode  occurs if  $\omega_r > |\omega_i|$; i.e., the amplitude of the 
coupled g-mode in the core is always exceedingly large.
The difference comes from the absence of an evanescent zone in the case of an
oscillatory convection mode, while in the p- and g-mode coupling there is a narrow
evanescent zone that separates the p-mode propagating envelope from 
the g-mode propagating core.
We note that the isolation of the oscillatory convective mode in the envelope 
does occur in helium shell flashing (thermal pulsing) models, 
in which a shell convection zone is present associated with the helium burning shell.
The shell convection zone prevents the g-mode oscillation from propagating into 
the core just as in the blue super-giants SPB stars \citep{sai06,god09}.
We have not included such flashing models in this paper 
because the short time duration of shell flashing 
means such stars will be rare.

\section{Comparison with observed LSP variability} \label{sec:discussion}
\subsection{Period-luminosity relation}

\begin{figure*}
\begin{center}
\epsfig{file=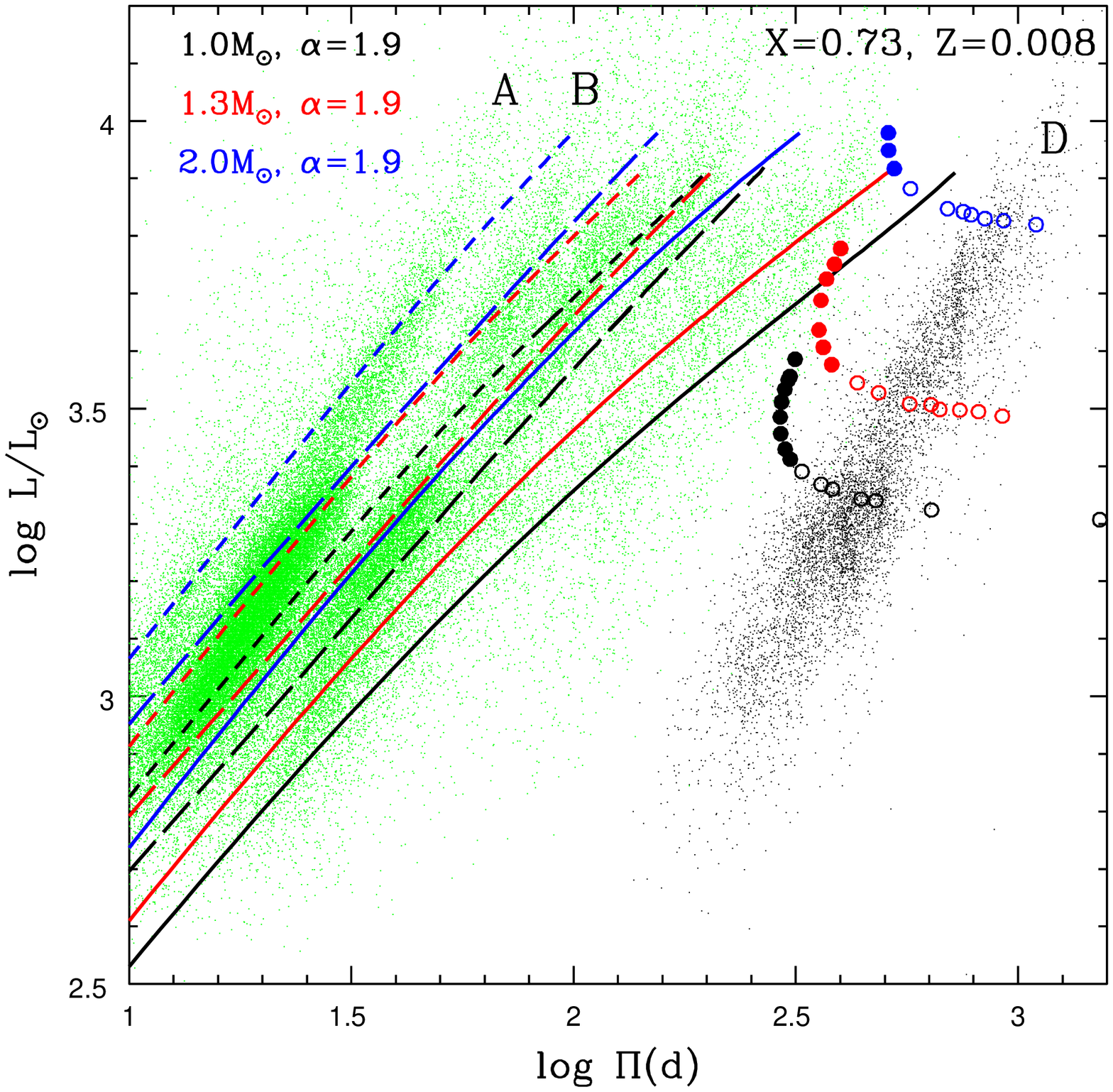,width=0.49\textwidth}
\epsfig{file=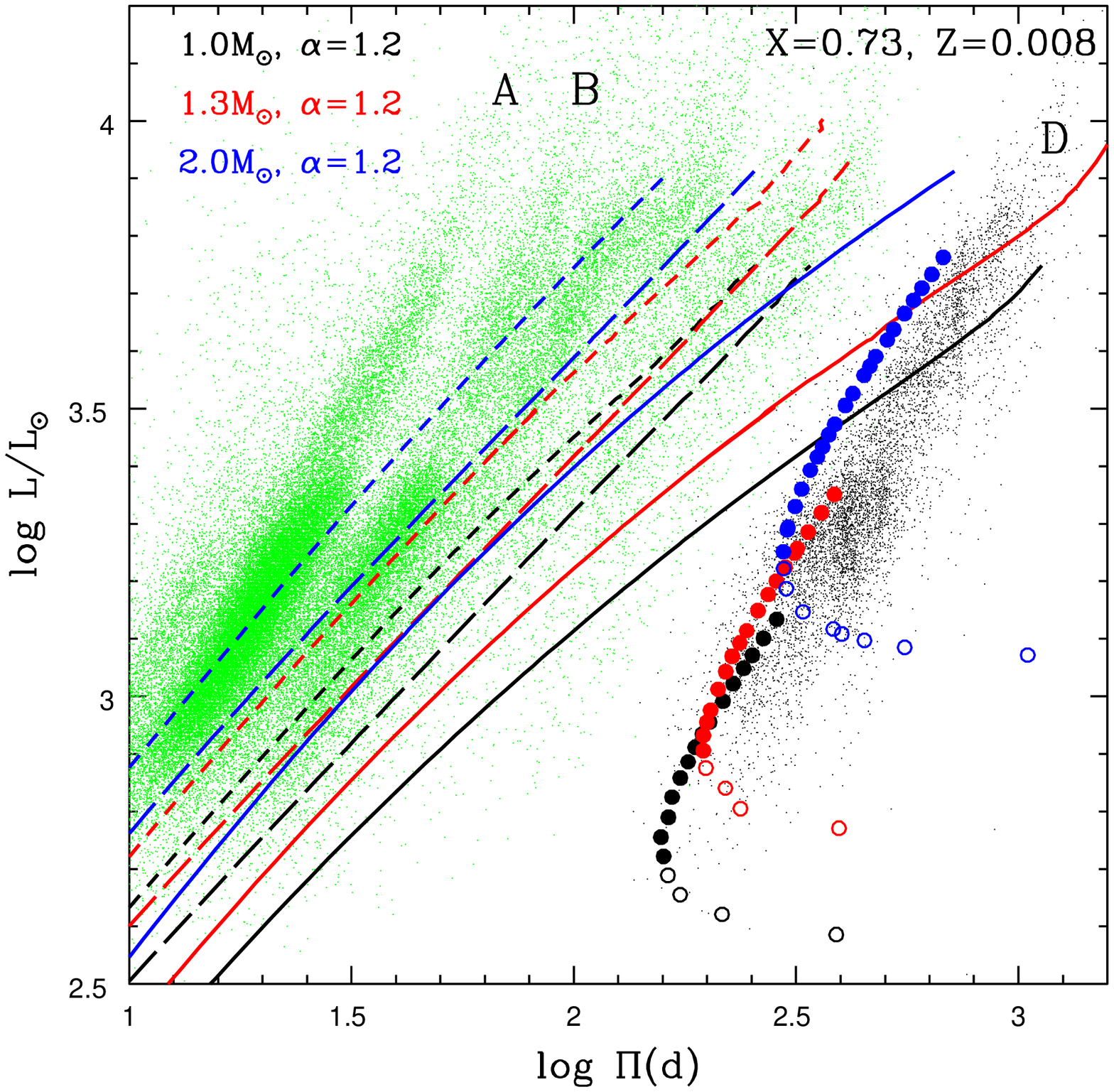,width=0.49\textwidth}
\end{center}
\caption{
Period-luminosity relations of dipole oscillatory convective modes (circles) 
and radial modes in AGB models (non-flashing phases) with mixing length 
parameters of  1.9 (left)  and 1.2 (right).
Filled circles are for the modes with
$\omega_{\rm r} > 0.5|\omega_{\rm i}|$ or vise versa for open circles.
Solid lines, long dashed lines, and short dashed lines correspond to the radial fundamental, 
first overtone, and second overtone modes, respectively. 
A maximum luminosity of each period-luminosity relation for dipole oscillatory convective modes
arises because in more 
luminous models coupling with a high-order g-mode in the core is significant and hence
the amplitude in the core is extremely high. 
Also shown are red-giant variables in the LMC with LSPs on sequence D (small black dots) 
and pulsating red-giant variables with periods on sequences C, C$^\prime$, B and A
(small green dots).
The periods come from OGLE III \citep{sos09}.
}
\label{fig:PL}
\end{figure*}

Fig.~\ref{fig:PL} compares period luminosity (PL) relations of red giants in the LMC with theoretical period-luminosity relations
of oscillatory convection modes (circles) and radial pulsations (solid and dashed lines). 
Models with  $\alpha = 1.9$ are shown in the left panel and models with  $\alpha = 1.2$ are shown in the right panel.
Our aim here is to match the periods of oscillatory convection modes with those of the LSPs belonging 
to sequence D while simultaneously matching the radial pulsation modes to sequence C (fundamental radial mode)
and the shorter-period pulsation sequences.
Period data \citep{sos09} for red-giant variables were taken from
the OGLE-III website (http://ogle.astrouw.edu.pl/).  
The luminosity of each star was obtained in the same way as for the sequence D stars 
shown in Fig.~\ref{fig:hrd}.

The pulsation period is proportional to $\sigma_r^{-1} \propto R^{1.5}/\omega_r$
(eq.~\ref{eq:omega}). 
The PL relations for radial pulsations are roughly straight lines because
$\omega_r$ of each mode does not vary much along the AGB evolution.
In contrast,  the PL relation of  the oscillatory convective mode has
a peculiar shape.
It bends and becomes nearly horizontal in less luminous parts, having
a minimum period and a minimum luminosity.
(The maximum luminosity corresponds to the point where $\omega_r = |\omega_i|$
as discussed in the previous section.)
The bending in the PL relation is caused by a rapid change in $\omega_r$ 
as a function of luminosity near the minimum luminosity  
(see Fig.~\ref{fig:lum_omega}). 
Around the minimum luminosity, small $\omega_r$ governs the PL relation; 
period decreases rapidly as luminosity increases due to an increase in $\omega_r$.
At a certain luminosity, however, the $R^{1.5}$ effect exceeds the $\omega_r$ effect
so that the period starts to increase with luminosity ($R$).

For models with the mixing-length parameter $\alpha = 1.2$ 
(right panel of Fig.~\ref{fig:PL}), the PL relation of the oscillatory
convective modes above the `bend' is 
consistent with sequence D i.e.,  periods, the gradient of the relation, and 
the lower luminosity bound of sequence D agree with models of 
$M \ga 1.0~{\rm M}_\odot$.
However, these models are cooler than the red-giant sequence of the LMC
in the HR diagram as seen in Fig.~\ref{fig:hrd} and the
radial mode periods, at least for masses less than $\sim 2 {\rm M}_{\odot}$,
are longer than the observed periods.

For models with $\alpha = 1.9$ (left panel of Fig.~\ref{fig:PL}), 
which best agree with the distribution of 
red giants on the HR diagram (Fig.~\ref{fig:hrd}) in the LMC
and which provide a reasonable match between 
radial-mode periods and observed pulsation periods, 
the periods of oscillatory convective modes above the `bend'
are too short for sequence D by a factor of  $\sim 2$.
Periods can cross sequence D only below the `bend',
for which $\omega_r < 0.5\omega_i$ (open circles).
Also, models of $M \ge 1.0~{\rm M}_\odot$ cannot explain the lowest part
of the sequence D;
the minimum luminosity for $1.0~M_\odot$ is $\log L/L_\odot \approx 3.3$,  
which is higher than the lower bound of the sequence D ($\log L/L_\odot \approx 2.8$).
Although the minimum luminosity for $\sim0.8~M_\odot$ models decreases as low
as the lower bound of the sequence D, the ages of these models exceed the cosmic age.

The problems in the $\alpha = 1.9$ models are caused, at least partially,
by the weak nonadiabaticity. 
The strength of nonadiabatic effects in a convection zone depends on the distribution 
of density and temperature, which in turn depends on the convection model. 
In this paper, we have  
used the mixing-length theory, which is a poor approximation for the convection in 
luminous red-giants, and we have neglected turbulent pressure and possible overshooting. 
A better convection theory including turbulent pressure and possible overshooting 
might increase  nonadiabatic effects  significantly in outer layers,  which could
extend the luminosity range in which the oscillatory convective modes
are found.

\subsection{Phase relations}

\begin{figure*}
\begin{center}
\epsfig{file=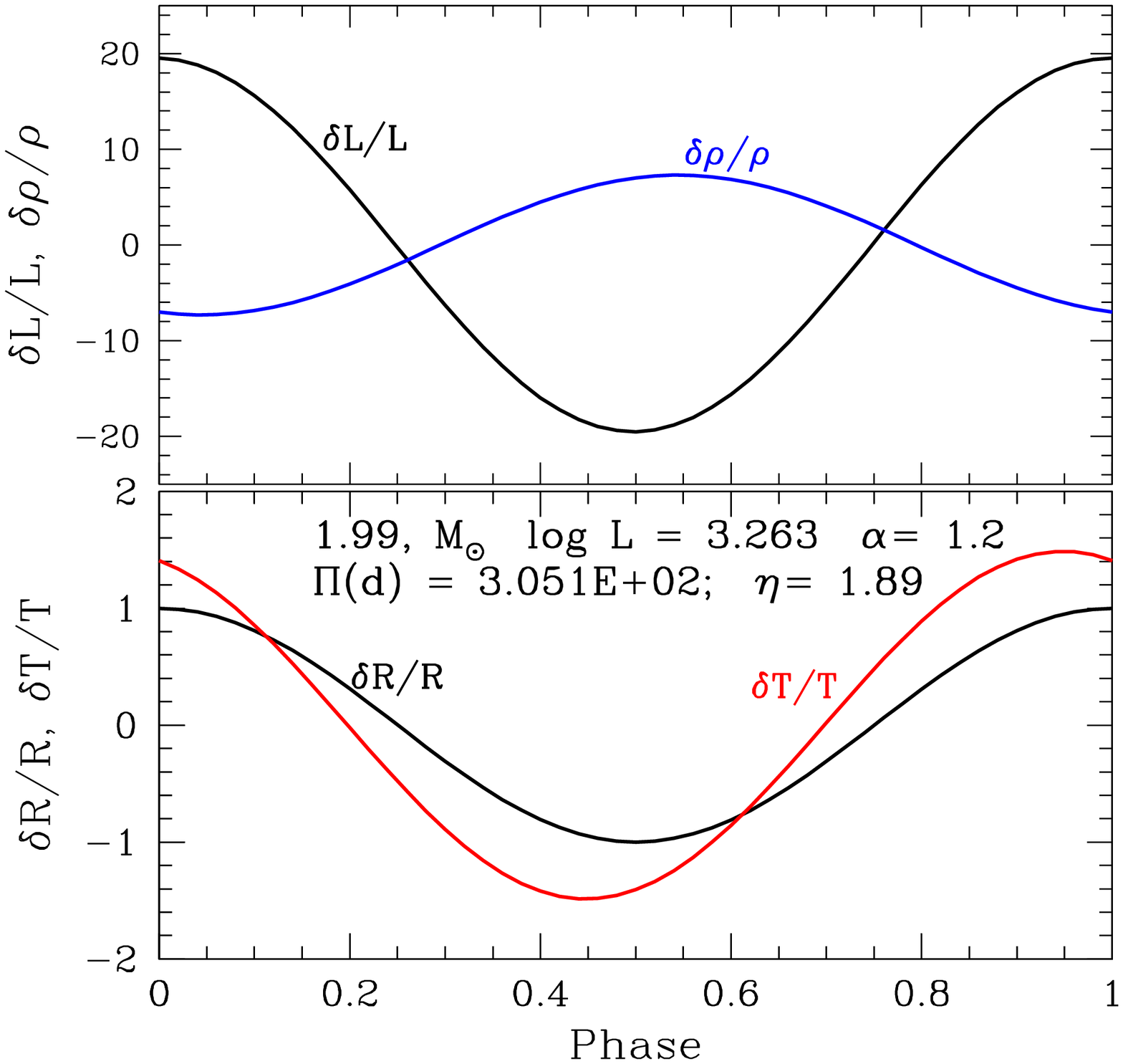,width=0.33\textwidth}
\epsfig{file=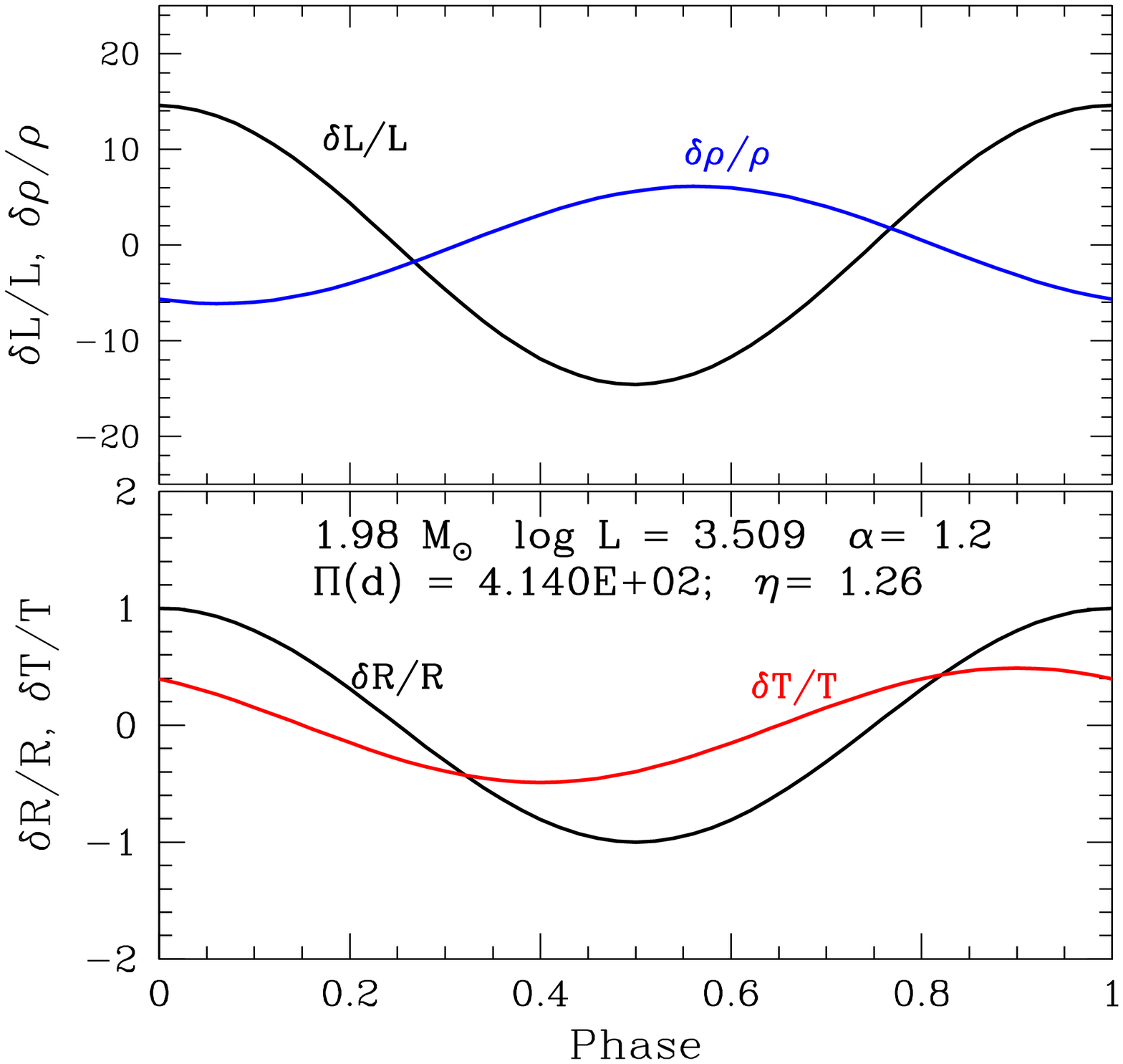,width=0.33\textwidth}
\epsfig{file=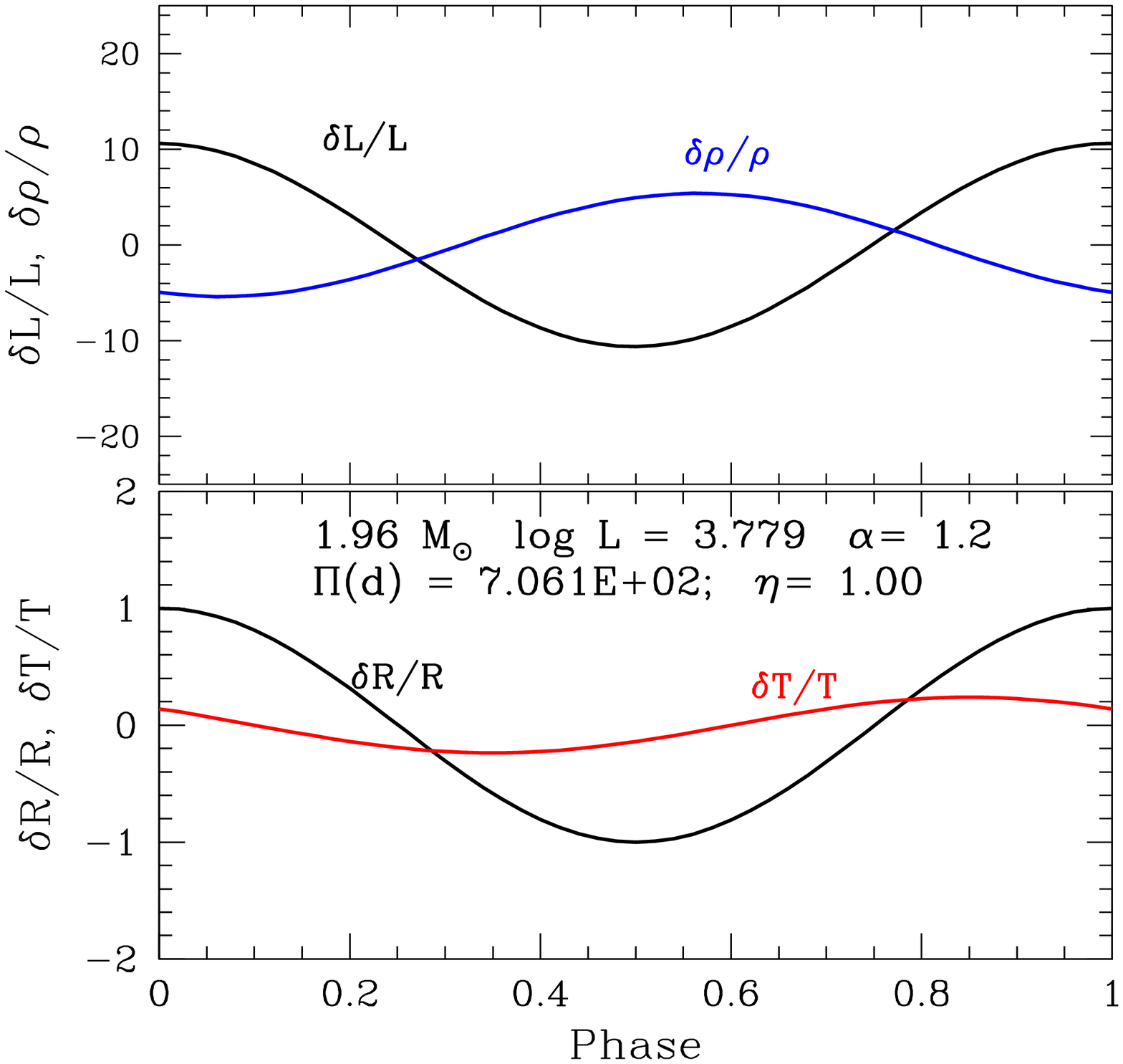,width=0.33\textwidth}
\end{center}
\caption{One cycle of  the photospheric temperature variations,$\delta T/T$, radial displacements, $\delta R/R$ (bottom panels), and photospheric 
density variation, $\delta\rho/\rho$, 
and luminosity variation at the outer boundary,
$\delta L/L$ (top panels) as a function of pulsation phase for dipole oscillatory convection modes
in three models of $2.0~{\rm M}_\odot$ ($\alpha = 1.2$) with different luminosities,
where $\Pi$(d) is period in days, and $\eta$ is the growth rate defined as 
$-\omega_i/\omega_r$. 
The left panel is for a model around the minimum period 
(Fig.~\ref{fig:PL}),  the middle panel is for the same model 
as in Fig.~\ref{fig:eigfunL1L2}, and the right panel is for a model close to the top of the 
period-luminosity relation in Fig.~\ref{fig:PL}.
The amplitude is normalizes as $\delta R/R =1$ at zero-phase, 
and is  proportional to a spherical harmonic $Y_1^m(\theta,\phi)$ with 
$\theta$ and $\phi$ being co-latitude and azimuthal angle.}
\label{fig:surf}
\end{figure*}

Fig.~\ref{fig:surf} shows predicted variations in luminosity, 
and photospheric density (top panels),
temperature and radial displacements (bottom panels) 
as functions of pulsation phase of dipole ($\ell = 1$) modes for
three $2~{\rm M}_\odot$ models with different equilibrium luminosities;
around the minimum period, middle, and the top of the PL sequence.
The photospheric variations of $\delta T/T$, $\delta\rho/\rho$ and
$\delta R/R$ were calculated using the eigenfunction at the mass shell
corresponding to the photosphere in the equilibrium model, while $\delta L/L$
stands for the luminosity variation at the outer boundary which is located at
an optical depth of $10^{-4}$.
We note that the quantities
shown in Fig.~\ref{fig:surf} are not averaged across the surface but
are local values proportional to the spherical harmonic
$Y_\ell^m(\theta,\phi)$.
In particular, a dipole mode does not change the spherical shape and the mean radius
of the stellar surface, although  the local distance from the center of mass varies
during pulsation. 
For this reason, we do not relate the observed radial velocity variation with 
a variation in the mean radius.

Fig.~\ref{fig:surf} shows that
temperature is higher when density perturbation is negative (positive buoyancy)
and radial displacement is positive for oscillatory convection modes. 
These phase relations are similar to those for normal convective instability, but 
are quite different from ordinary p- and g-modes.
Because of the peculiar phase relations in oscillatory convective modes, 
temperature and luminosity are higher when radial displacement is positive.
The amplitude of $\delta T/T$ relative to the radial displacement decreases
with luminosity; the ratio is about $1.5$ in a model at
$\log L/{\rm L}_\odot = 3.26$ (around the minimum period), while 
the ratio is reduced to about $0.2$ at $\log L/{\rm L}_\odot = 3.78$ 
(around the top of the PL relation). 
Non-adiabatic effects (thermal diffusion), which are stronger in more luminous models,
reduce temperature variations.
In addition, because of the stronger nonadiabatic effect, the phase relation between 
$\delta T/T$ and $\delta\rho/\rho$ shifts and the growth rate $\eta$ 
decreases as the luminosity increases.

\begin{figure}
\epsfig{file=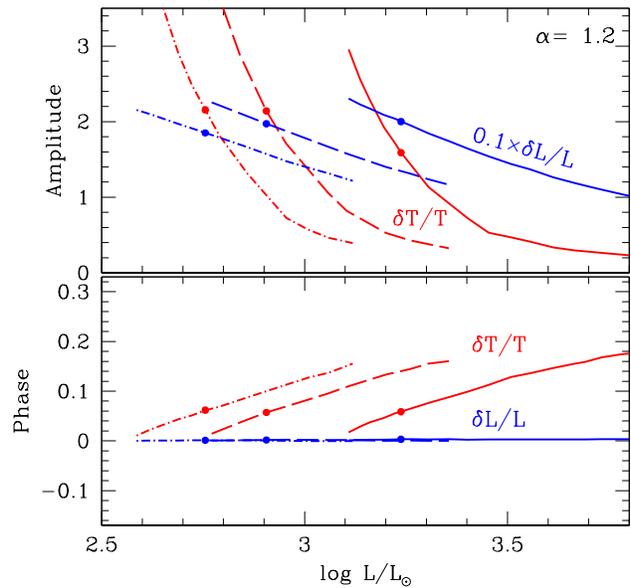,width=0.49\textwidth}
\caption{
Amplitude (upper panel) and phase (lower panel) for the  luminosity 
at the outer boundary (blue) and the  photospheric 
temperature (red) variations of oscillatory convective modes ($\ell=1$)  versus luminosity.
Solid, dashed, and dot-dashed lines are for $2.0, 1.3,$ and $1.0~M_\odot$ models, respectively. 
The amplitudes are normalized by the radial displacement, $\delta R/R$.
The phase is in units of period, with zero phase corresponding to the phase of maximum $\delta R/R$.
A filled circle indicates the position of the minimum period for each case.
}
\label{fig:amp_phase_L}
\end{figure}

Fig.~\ref{fig:amp_phase_L} shows systematic trends of the amplitude and 
phase of $\delta T/T$ and $\delta L/L$ as a function of luminosity for 
models of mass 2.0 (solid line), 1.3 (dashed line) and $1.0~M_\odot$ (dot-dashed line).
The behaviour relative to the luminosity at minimum period (filled circles)
is similar among the cases with different masses.
For a given mass, as the luminosity increases, the amplitudes of 
$\delta T/T$ and $\delta L/L$ decrease (relative to the amplitude of $\delta R/R$). 
The phase of $\delta T/T$ increases gradually with luminosity, 
while the phase of $\delta L/L$ stays very small (i.e. $\delta L$ is in phase with $\delta R$).
The phase difference between $\delta T/T$ (at photosphere) and $\delta L/L$ (at the outer boundary) arises because of  the entropy variation between 
the photosphere and the outer boundary.
The phase of the temperature variation at the outer boundary is very close to that of $\delta L/L$. 

\citet{nic09} found various statistical properties of the LSP variability. They
found that radius and temperature variations based on the radial
velocity variations are very different from those determined from
photometry or spectroscopic analysis (which would be expected if 
the pulsations are dipole modes). 
Here we summarize the properties
of the majority of their sample based on photometric and spectroscopic
analyses (i.e. we do not consider velocity data) as follows: (1) the phase delay of minimum light from the phase
of minimum radius is mostly less than 0.2 period, (2) $\Delta T_{\rm
  eff}/T_{\rm eff} \la 0.03$ and $\Delta R/R \sim 0.05$.
\footnote{We note that although the whole surface averages of $\Delta T_{\rm eff}$ and
$\Delta R/R$ should be zero for nonradial pulsations, 
observational values should be non-zero
because they correspond to the means across the visible hemisphere;
the cancellation effect is smallest in dipole modes among nonradial pulsations.}

In the oscillatory convective modes shown in Fig.~\ref{fig:surf},
minimum light occurs around minimum displacement which 
agrees reasonably well with the
property (1).  Because the pulsation is extremely nonadiabatic,
temperature variation is very small. 
The ratio of the amplitudes of temperature and radius variations in the
oscillatory convective mode ranges from about $2$ to $0.2$ (
from around minimum period to the top of the sequence; see  Figs.~\ref{fig:surf} and
\ref{fig:amp_phase_L}),
while the property (2) above indicates the ratio is $\la 0.6$ roughly
consistent with the theoretical prediction.  
We note that if we use only photometric data, which cover all pulsation phases, 
the average of the whole sample is found around $0.23$.   
Thus, the properties of
oscillatory convective modes seem consistent with the properties of
the majority of the LSP sample of \citet{nic09}.

Recently, \citet{tak15} examined color-magnitude variations during LSP
variations for some luminous SMC red giants. They found that matter
should be ejected near the beginning of light decline. The matter
might be related to the presence of chromosphere activity
\citep{woo04} and might be the origin of the excess mid-IR emission
observed around stars with LSPs \citep{woo09}.  Fig.~\ref{fig:surf}
indicates that maximum light occurs around maximum radial
displacement, where mass ejection might be expected to occur.
This is in agreement with phase of ejection found by \citet{tak15}.

\section{Conclusion}

We discussed  properties of the oscillatory convective modes 
in luminous red giants and compared them with the properties 
of the LSPs of variable red giants on the sequence D in the LMC.  

The phase relations among temperature, radial displacement, 
and density variations in the oscillatory convective mode 
are similar to the relation expected for convective eddies 
rather than those of ordinary p- or g-modes; i.e., a positive temperature
variation is associated with positive radial displacement and negative density
variation.
In addition, because of strong nonadiabatic effects the amplitude of the surface
temperature variations is reduced relative to the radial displacement.
These properties of the oscillatory convective mode roughly agree with the
properties of the LSP variations of red giants in the LMC.

The period-luminosity relation of the oscillatory convective modes 
is found to be consistent with 
the observed relation of the LSPs of red-giants along sequence D 
if we adopt a mixing-length as small as 1.2 times pressure scale-height ($\alpha=1.2$). 
However, these models are much cooler than the AGB stars in the LMC. 
For models with $\alpha = 1.9$, 
which agree with the effective temperatures of red giants in the LMC,
periods are too short and the lower bound of the luminosity is too high, 
inconsistent with sequence D.
Relatively weak nonadiabatic effect in the models with $\alpha=1.9$ would 
cause, at least partially, the discrepancy.
The shortcomings of the present models might be improved by employing a more sophisticated convection 
theory including turbulent pressure and possible overshooting. 
We leave the problem for future investigations.

\section*{acknowledgements}
We thank Bill Paxton and the MESA project team for developing the efficient stellar evolution code MESA.  
 We also thank Marc-Antoine Dupret for constructive comments and suggestions as the referee of this paper.

\bibliography{seqD}



\label{lastpage}

\end{document}